\documentclass[%
aip,
amsmath,amssymb,
]{revtex4-1}
\usepackage[utf8]{inputenc}
\usepackage{mathtools}
\usepackage{natbib}
\usepackage{graphics}
\usepackage{color}
\usepackage{mathrsfs}
\usepackage{floatrow}   
\usepackage{graphicx,xcolor}
\usepackage[framemethod=tikz]{mdframed}
\usepackage{amssymb,amsmath,amsfonts,amstext}
\usepackage{stmaryrd}
\usepackage{verbatim}
\usepackage{amsthm}
\usepackage{esint}
\usepackage{caption}
\usepackage{subcaption}
\usepackage{bm}
\graphicspath{ {./Figures/} }

\makeatletter
\def\@email#1#2{%
 \endgroup
 \patchcmd{\titleblock@produce}
  {\frontmatter@RRAPformat}
  {\frontmatter@RRAPformat{\produce@RRAP{*#1\href{mailto:#2}{#2}}}\frontmatter@RRAPformat}
  {}{}
}%
\makeatother
\begin{document}

\title{Exploring the Relationship Between Softness and Excess Entropy in Glass-forming Systems}
\author{Ian R. Graham}
    \affiliation{Department of Physics and Astronomy, University of Pennsylvania, Philadelphia, PA 19104, USA}
\author{Paulo E. Arratia}
    \affiliation{Department of Mechanical Engineering and Applied Mechanics, University of Pennsylvania Philadelphia, PA 19104, USA}
\author{Robert A. Riggleman}
    \affiliation{Department of Chemical and Biological Engineering, University of Pennsylvania Philadelphia, PA 19104, USA}
\date{\today}

\begin{abstract}
 We explore the relationship between a machine-learned structural quantity (\textit{softness}) and excess entropy in simulations of supercooled liquids. Excess entropy is known to scale well the dynamical properties of liquids, but this quasi-universal scaling is known to breakdown in the supercooled and glassy regimes. Using numerical simulations, we test whether a local form of the excess entropy can lead to predictions that derive from softness, which has been shown to correlate well with the tendency for individual particles to rearrange. To that end, we explore leveraging softness to compute excess entropy in the traditional fashion over softness groupings. Our results show that by computing the excess entropy over softness-binned groupings, we can build a strong quantitative relationship between the rearrangement barriers across the explored systems.

\end{abstract}

\maketitle

\section{Introduction}

Despite their importance to numerous technologies, the properties of disordered (amorphous) materials far from equilibrium continue to elude comprehensive understanding \citep{stillinger_glass_2013, charbonneau_glass_2017, berthier_theoretical_2011}. Disordered materials made of particulates can span a wide range of length scales, and examples include metallic glasses, nanoparticle packings, colloidal suspensions, foams \& emulsions, and granular materials \cite{cubuk_structure-property_2017, hecke09, jerol19}. Our knowledge of these materials is primitive compared to our understanding of crystalline solids, where symmetry and order guide detailed theories \cite{torquato2018, LiuNagel_ARCM_2010}. In disordered systems, the dynamics are known to be strongly heterogeneous and can vary by orders of magnitude even for a supercooled liquids at equilibrium. For many glass-forming systems, the dynamics exhibit a temperature dependence where the activation energy grows upon cooling. Thus, relating the evolution of the material microstructure (at the constituent level) to the material's dynamics (and bulk response) is quite difficult. Much work has been devoted to developing structural indicators of glassy dynamics in disordered systems, with different degrees of success \citep{richard_predicting_2020, galloway_scaling_2020, dyre_perspective_2018, bonnecaze_excess_2020}. If one is interested in developing thermodynamically consistent coarse-grained and constitutive models, identifying the relevant structural parameters and their connection to thermodynamic quantities is a critical step \citep{ottinger_beyond_2005, ottinger_systematic_2007}.

In recent years, machine learning (ML) and data science techniques have matured to a point that they are ubiquitous in research and industry. Past studies have successfully applied ML models to predict local dynamical properties of glassy systems more accurately than traditional quantities such as local potential energy and free volume, for example \citep{tong_order_2014, riggleman_tuning_2007, widmer-cooper_free_2006, richard_predicting_2020}. These ML models range in complexity from relatively simple linear support vector machines (SVMs) which operate on vectors of pre-selected structural features for each particle, to convolutional and graph models that are able to extract more intricate spatial correlations within the material and thus lead to improved prediction accuracy on similar tests \citep{schoenholz_structural_2016, schoenholz_relationship_2017, bapst_unveiling_2020, fan_predicting_2021}. A notable ML indicator developed to characterize structural defects and predict rearrangements in disordered packings is the quantity known as \emph{softness} \cite{schoenholz_structural_2016,cubuk_structural_2016, cubuk_structure-property_2017}. This quantity was one of the first models developed for glasses and, though relatively simple in its construction from local radial distribution functions, is still able to provide good predictive power \citep{schoenholz_structural_2016, schoenholz_relationship_2017}.

Although \textit{softness} may be more interpretable compared to other machine-learning based models, particles of similar softness still possess different local structures, complicating its overall physical meaning. For systems in which particles are dominated by isotropic interactions, \textit{softness} can be constructed as a weighted sum of the radial distribution function, where the weights are determined by the SVM. While these weights do inform us to a degree that certain features in the pair correlation function (notably the first and second peaks of $g(r)$) are critical to local dynamics, it remains unknown how the general form of the radial weights may be related to \textit{a priori} knowledge of the particles and their interactions. On the other hand, excess entropy, a quantity known to scale with the dynamical properties of simple liquids, can be similarly constructed as a simple function of local particle coordinates where deviations in the radial distribution function from that of the ideal gas (a flat distribution) are penalized \citep{dyre_perspective_2018, bell_excess-entropy_2020, bonnecaze_excess_2020, ma_excess_2019, galloway_scaling_2020}. Due to the similar construction of these quantities, it is tempting to consider whether there are any connections between them. This would lead to an improved understanding of \textit{softness}, for example. However, one difficulty of such a comparison is that \textit{softness} is defined on a particle basis to strongly correlate with particle-level dynamics, while excess entropy is usually defined for an entire configuration and is associated with system-average dynamics. In this work, we formulate a comparison between these quantities (i.e., excess entropy and \textit{softness}) using two separate approaches. First, we define a local form of excess entropy that takes in a Gaussian-smeared, coarse radial distribution function (similar to \textit{softness}) and compare the performance of this measure relative to \textit{softness}. Next, we utilize \textit{softness} as an intermediary tool so to compute excess entropy in a more traditional fashion. Essentially, we treat particles of similar \textit{softness} as comprising their own ensembles, allowing us to compute ensemble average quantities (like excess entropy) over \textit{softness}-grouped subsystems. Our results suggest interesting future directions for combining equilibrium tools with machine-learned quantities in out-of-equilibrium disordered systems.

\section{Methods}

\subsection{Simulation Details}

Numerical simulations comprise sets of equilibrated supercooled states of bidisperse, Lennard-Jones character generated using HOOMD-blue \citep{anderson_hoomd-blue_2020}. Configurations are composed of 32,768 particles in a 3D periodic box and a Nos\'e-Hoover thermostat is used to integrate the dynamics with a time step of $10^{-3}$ in the NVT ensemble. We employ a Lennard-Jones potential with a modifiable well-width parameter $\Delta$ to control the level of caging in the system; the standard Lennard-Jones definition is obtained when $\Delta=0.0$. The potential is defined as 
\begin{equation} \label{Eq:LJPot}
\begin{split}
    \cr 
    V_{ij}(r) = \begin{cases} 
        4\epsilon_{ij}\left[\left(\frac{\sigma'_{ij}}{r-\Delta}\right)^{12} - \left(\frac{\sigma'_{ij}}{r-\Delta}\right)^6\right] & r\leq 2.5\sigma_{ij} \\
        0 & r > 2.5\sigma_{ij},
    \end{cases}
\end{split}
\end{equation}
where $\sigma'_{ij} = \sigma_{ij} (1-\Delta/{2^{1/6}})$ is defined to keep the minimum at the same position as $\Delta$ varies, $r$ is the pair distance between particles, $\epsilon_{ij}$ is the interaction energy scale between species $i$ and $j$, and $\sigma_{ij}$ is the interaction length scale. The parameter $\Delta$ is varied from 0.0 to 0.4, and $\epsilon_{ij}$ and $\sigma_{ij}$ are set in accordance with a standard 80:20 Kob-Anderson type mixture ($\epsilon_{AA}=1.0$, $\epsilon_{AB}=1.5$, $\epsilon_{BB}=0.5$, $\sigma_{AA}=1.0$, $\sigma_{AB}=0.8$, $\sigma_{BB}=0.88$). The functional forms are shown in Figure \ref{Fig:Potential}a for different choices of $\Delta$. Depending upon the value of $\Delta$ used, packing fraction $\rho$ is varied between 1.2 and 1.12 to minimize the rapid increase in system pressure $p$ as the well width is decreased. We explored the $(\Delta, \rho)$ pairings [$(0.0, 1.2)$, $(0.1, 1.18)$, $(0.2, 1.16)$, $(0.3, 1.14)$, $(0.4, 1.12)$] . Simulations at $\Delta=0.5$ were also produced, but they were found to readily crystallize in the supercooled regime; they are omitted from the analysis. To ensure reproducibility, three randomly seeded replicas were generated for each system. All stated Lennard-Jones units are with respect to the A-species of the standard Kob-Anderson mixture (i.e., $\Delta=0$). Configurations are initialized on a cubic lattice with random placements of particle species and thermalized well into the liquid regime for each system. The systems are then swept through the analyzed temperatures into the supercooled regime, waiting at most 20 $\tau_\alpha$ (where $\tau_\alpha$ is the alpha relaxation time) at the highest temperature in the supercooled regime and 8 $\tau_\alpha$ for the coldest temperature before continuing with the temperate quench. We use the self-intermediate scattering function ($F(\bm{Q},t)=\frac{1}{N}\sum_{j=1}^N\langle\exp[i\bm{Q}\cdot(\bm{r}_j(0)-\bm{r}_j(t))]\rangle$) to estimate $\tau_\alpha$ by calculating the time required for $F(\bm{Q},t)$ to drop below $\frac{1}{e}$. In all systems we use $|\bm{Q}|=7.14$, the wavenumber coinciding with the distance to the first peak of $g(r)$. Furthermore, we apply FIRE minimization as a post-processing step to obtain the inherent structures at each sampled time for final analysis; this step is not strictly necessary to obtain the relationships we find, but serves to remove thermal fluctuations from the analysis \citep{bitzek_structural_2006}. In all parts of the analysis, we explore only the dynamics of A-species particles. We made substantial use of the \emph{freud}, \emph{signac}, and \emph{signac-flow} python packages to perform post-processing analysis and to manage data \& job workflows \citep{ramasubramani_freud_2020, ramasubramani_signac_2018, adorf_simple_2018}.

\subsection{Softness}

As briefly discussed above, \textit{softness} ($S$) is a machine-learned quantity trained on examples of particles undergoing rearrangement. Here, we closely follow the techniques employed in previous works to construct $S$ in our thermal system \citep{schoenholz_relationship_2017, schoenholz_structural_2016}. We assess whether or not a particle participated in a rearrangement by examining $p_{hop}$. $p_{hop}$ is defined as
\begin{equation}
    p_{hop}(i,t) = \sqrt{\langle(x_i - \langle x_i \rangle_{B} )^2 \rangle_A \langle(x_i - \langle x_i \rangle_{A} )^2 \rangle_B} .
\end{equation}
where $x_i$ is the position of the particle $i$ in the simulation, $A$ and $B$ are time intervals defined as $A=[t-t_R/2,t]$ and $B=[t,t+t_R/2]$, and $t_R$ is the time window used. A fairly coarse period is used between frame dumps of $1\tau_A$, where $\tau_A$ is the Lennard Jones time unit in reference to the A species of the $\Delta=0.0$ system, and we use a time window $t_R=10\tau_A$. To categorize rearrangements, two cutoffs are defined: $p_{H}=0.05$ and $p_{S}=0.2$. Particles are deemed \emph{soft} if $p_{hop}>p_{S}$ and \emph{hard} if $p_{hop}<p_{H}$. We further restrict our dataset by only processing the peaks and troughs of $p_{hop}$, and asserting that rearrangement events are separated by $10\tau_A$, and non-rearrangements by $80\tau_A$. In case any events are too close, the more prominent peak (or deepest trough) is selected.

Structure functions, $\mathcal{G}_K(\mu_{i})$, are calculated from a Gaussian-smeared radial distribution function by the following equation 
\begin{equation}
    \mathcal{G}_{K}(\mu_i) = \sum_{j \in \{K\}}{e^{\frac{(r_j-\mu_i)^2}{2{\Delta \mu}^2}}}
\end{equation}
where $\mu_i$ is the radial distance of the $i^{\textrm{th}}$ structure function, $\Delta \mu$ is variance of the gaussian, the label $K$ selects for particle species A or B, and $r_j$ is the radial distance to the particle $j$. $\mu_i$ varies linearly from 0.4 to 3.0 with a step size of $\Delta \mu = 0.1$ \citep{behler_generalized_2007}. With these structure functions, we proceed to train a LinearSVM (using the \emph{scikit-learn} package\citep{pedregosa_scikit-learn_nodate}) to classify our soft and hard particles. The final hyperplanes are insensitive to the random seeds relating to the SVM optimization and subset of data used. The LinearSVM works by finding an optimal hyperplane to separate our classes in the high-dimensional space in which our structure functions reside. Note that we apply a pre-processing step to our data that shifts and scales the structure functions such that within each index pair ($K,\mu_i$) our data has zero mean and unit variance. After fitting the data to our model, we can extract the decision function which is the signed distance to the hyperplane and use this measure as our softness $S$. We find that the accuracy, measured as $P(R|S>0)$, is between 73-80\% for the coldest temperatures measured in each system, though this accuracy decreases to 40-50\% at our highest temperatures due to the inherent thermal stocasticity of the liquid state.

\subsection{Excess Entropy}

Excess entropy, $s^{(2)}$, is classically defined as the difference in configurational entropy between the ideal gas and the system of interest \citep{dyre_perspective_2018}. It can be calculated by a variety of means, including thermodynamic integration \citep{bell_excess-entropy_2020}. Here, we utilize the simplified and approximate calculation based upon the species-dependent radial distribution function $g_K(r)$ as
\begin{equation}
s^{(2)}=-2 \pi \sum_K \rho_K \int ( g_K(r)\log\{g_K(r)\} - g_K(r) + 1)r^2 dr
\end{equation}
in 3D. $\rho_K$ is the density of species K in the sample and $r$ is the radial distance. We additionally construct a per-particle version of $s^{(2)}$ in close analogy to  \textit{softness}. The local $s^{(2)}$ utilizes the same structure functions as \textit{softness}, but these inputs are rescaled by the appropriate spherical measure to be transformed back into a coarse, smeared $g_K(r_i)$.  
\begin{equation}
    g_K(r_i) = (4 \pi \rho_K^* r^2_i)^{-1}\mathcal{G}_{K}(r_i)
\end{equation}
where $\rho_K^*$ is an effective local density calculated by integrating the RDF for each species, here computed with a sphere of radius $r=3.0$. From this RDF, we can apply a simple midpoint integration to obtain 
\begin{equation}
s^{(2)}=-2 \pi \sum_K \rho_K^* \sum_i^{i_{max}}{[g(r_i)\log\{g(r_i)\} - g(r_i) + 1]r_i^2 L(r_i-\eta) \Delta r}
\end{equation}
where $\Delta r$ is the bin width, and $L(r_i-\eta)$ is the inverse logistic function centered at a distance of $\eta$ away. We applied a gentle optimization pass, finding $\eta=2.0$ to adequately dampen fluctuations at distances at and beyond the second peak of $g(r)$. A few aspects of the form we have chosen here substantially improve the performance of this measure. Notably, respecting the bidispersity of the systems and using the local species density $\rho_K^*$, as opposed to the average system density.

\section{Results}

\subsection{Comparison of softness and local excess entropy}

\begin{figure}[H]
    \centering
    \includegraphics[width=1.0\textwidth]{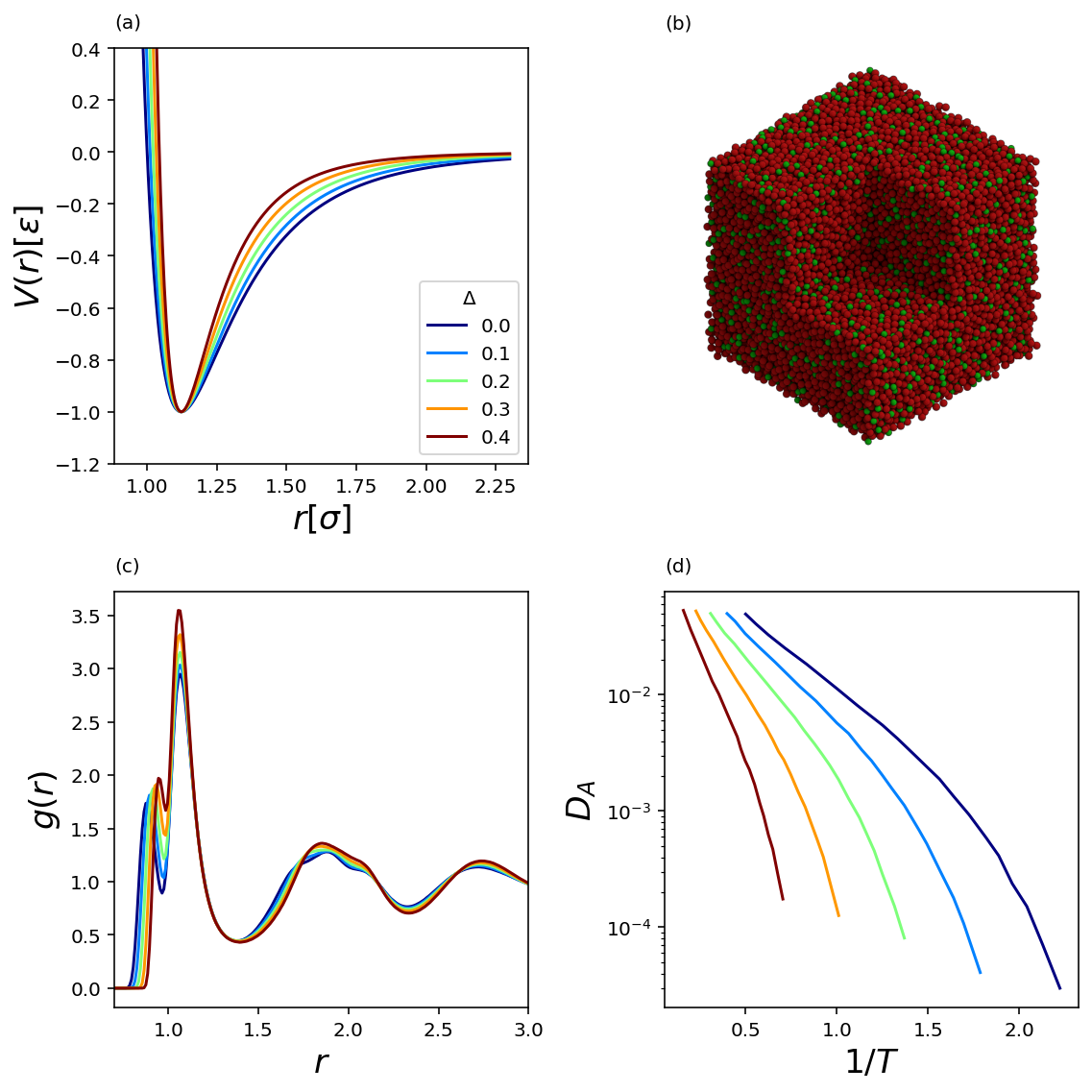}
    \caption{(a) Pair potentials used in this work. Marginal variation in the Lennard-Jones well width, controlled here by the parameter $\Delta$, leads to significant dynamical changes. (b) Example snapshot of our standard Lennard-Jones Kob-Anderson mixture at $k_B T = 0.45$, $\Delta=0.0$, produced with the \emph{fresnel} python library. (c) The total radial distribution functions (RDFs) for the 5 systems we consider. Higher $\Delta$ values lead to a sharpening of RDF peaks, in turn leading to more pronounced caging and slower dynamics at a given temperature. (d) Diffusion coefficients for A species particles measured in both the liquid and supercooled regimes.}
    \label{Fig:Potential}
\end{figure}

Here, we examine simulations of a Lennard-Jones Kob-Anderson mixture \citep{kob_scaling_1994}. We explore four other variations of this system with a modified well width parameter $\Delta$, ranging from 0.0 to 0.4, as shown in Figure \ref{Fig:Potential}a. Although the modifications to the potential appear subtle, this translates into considerably stronger caging effects within the system and more brittle response overall as $\Delta$ is increased \citep{lin_distinguishing_2019}. Figures \ref{Fig:Potential}b,c show an example snapshot of our $\Delta=0.0$ system at $k_B T = 0.45$ and samples of $g(r)$ at similar depths into the supercooled regime for each system. The increased caging behaviour is reflected in the increased height of the $g(r)$ peaks when $\Delta$ is increased. Figure \ref{Fig:Potential}d shows the diffusion coefficients for these five systems as a function of inverse temperature; particle diffusivity decreases considerably as $\Delta$ is increased for same $T$. 

The dynamic properties of liquids, such as diffusion, at high temperatures are expected to exhibit an Arrhenius temperature dependence, implying that a characteristic energy describes the temperature dependence of the diffusivity or viscosity. In this regime details of individual particle microstructure are unimportant, and all particles of the same type obey similar dynamics. At short time scales, particles travel ballistically as they move through their local neighborhood, but the short mean free path of the liquid ensures that these trajectories are quickly redirected, and diffusive behavior soon follows. As the system is cooled, it reaches a temperature where the Arrhenius trend begins to break down, known as the crossover temperature. For temperatures below the crossover, details of the individual particle environments become more important as particles spend more time in any given local configuration. This coincides with the emergence of stretched exponential behaviour in the self intermediate scattering and overlap functions, which is understood to result from an underlying wide distribution of relaxation rates within the system. This emerging importance of the specific microstructure, and resultant wide distribution of relaxation times, greatly complicates the prediction of dynamics in these systems. It becomes mostly hopeless to make first-principles predictions of these behaviours from information of the potential and macroscopic observables alone.

\begin{figure}[H]
    \centering
    \includegraphics[width=1.0\textwidth]{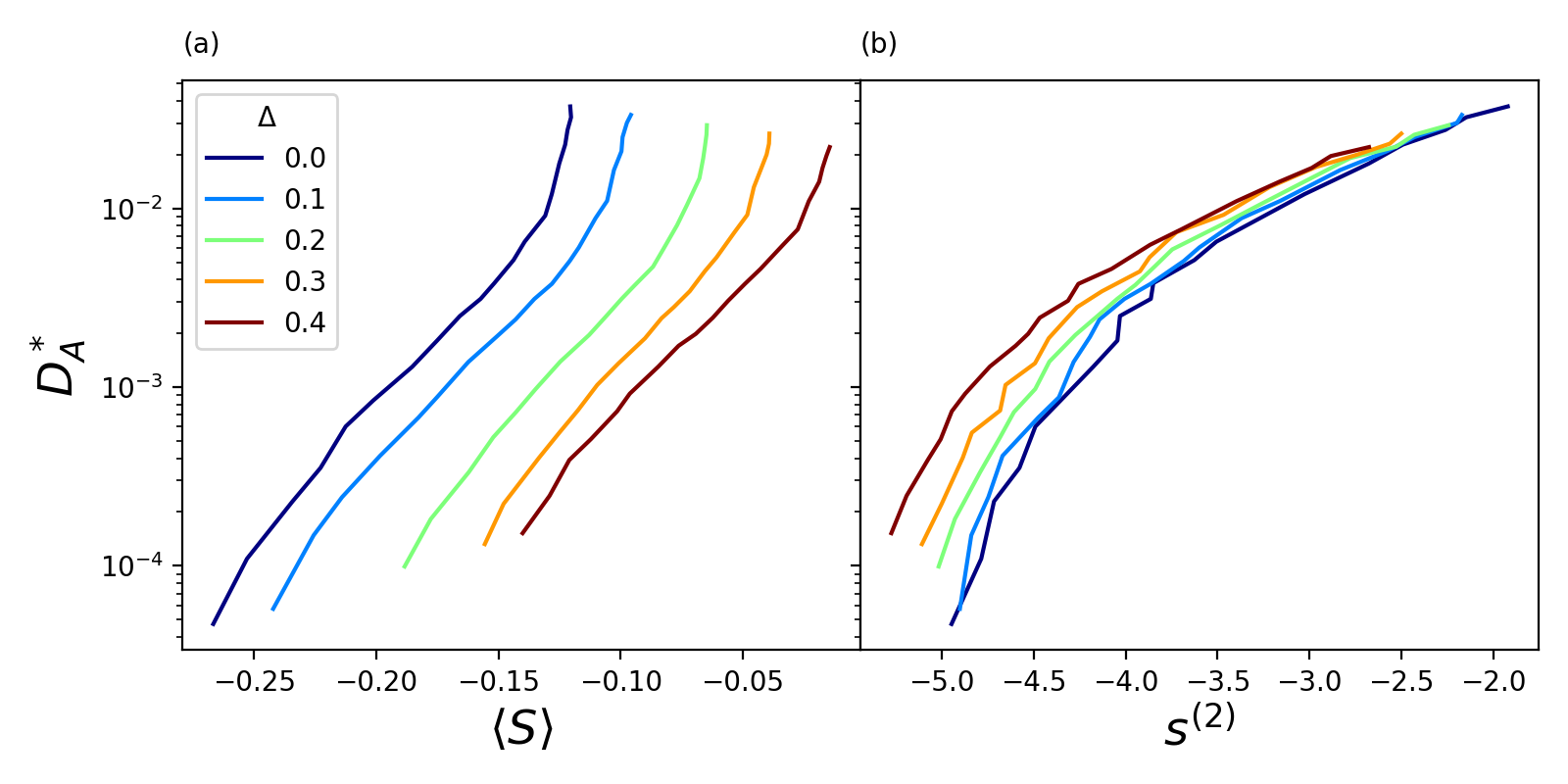}
    \caption{Reduced diffusion coefficients of the A species as a function of both the average softness (a) and excess entropy (b). $D_A^* = \rho^\frac{1}{3}\sqrt{\frac{m}{kT}}D_A$. The behaviour of these systems as softness varies, while qualitatively quite similar, fails to be agree quantitatively. The results regarding excess entropy though depend upon the regime. In the liquid state the excess entropy yields good agreement in the reduced diffusion coefficient across systems, though this breaks down considerably below the crossover temperature.}
    \label{Fig:ReducedDiffusionExEntr}
\end{figure}

In Figure \ref{Fig:ReducedDiffusionExEntr} we illustrate how the reduced diffusion coefficient, calculated as $D_A^\ast = \rho^\frac{1}{3}\sqrt{\frac{m}{kT}}D_A$, correlates with both the average softness ($S$) and excess entropy ($s^{(2)}$) of the systems. The prefactor of the reduced diffusion coefficient originates from Enskog theory, i.e. the kinetics of a dense hard-particle gas \citep{dzugutov_universal_1996}. As expected, we observe monotonic behavior in both quantities, but they differ qualitatively in both the liquid and supercooled regimes. In the liquid state, the average $S$ varies little. This is consistent with the understanding that \textit{softness} is well correlated with the effective energy barrier of particles, which should be constant in the liquid regime. As the temperature is decreased below the onset temperature and the average values of $S$ take on a more substantial temperature dependence where $S$ decreases with decreasing $T$, the relationship between $D_A^\ast$ and softness changes to a nearly exponential relationship, though the relation is clearly stronger than a simple exponential at the lowest temperatures considered. While these trends are qualitatively similar across the different systems, it remains difficult to relate them quantitatively due to the construction of $S$. This is because $S$, obtained here as a signed distance from an SVM hyperplane where all dimensions have been rescaled to unit variance over the data distribution, has poorly defined units. And it is a non-trivial issue how one would rescale and shift these dimensions such that particles of a given softness correspond to, for example, the same (reduced) diffusion coefficient from first principles.

For the case of excess entropy in Figure \ref{Fig:ReducedDiffusionExEntr}b, we see expected behaviour where in the liquid state the pair approximation of excess entropy does of fair job of estimating the reduced diffusion coefficients. However, the trends quantitatively differ once temperatures drop below onset. Application of the 3-body term in the excess entropy or its estimation using thermodynamic integration could perhaps improve the collapse in the liquid state, however we wish to limit our analysis to predictors that only require a single (or small number) of configurations. Unfortunately, both quantities, $S$ and $s^{(2)}$, fall short as universal predictors of bulk dynamics in the supercooled regime.

\begin{figure}[H]
    \centering
    \includegraphics[width=1.0\textwidth]{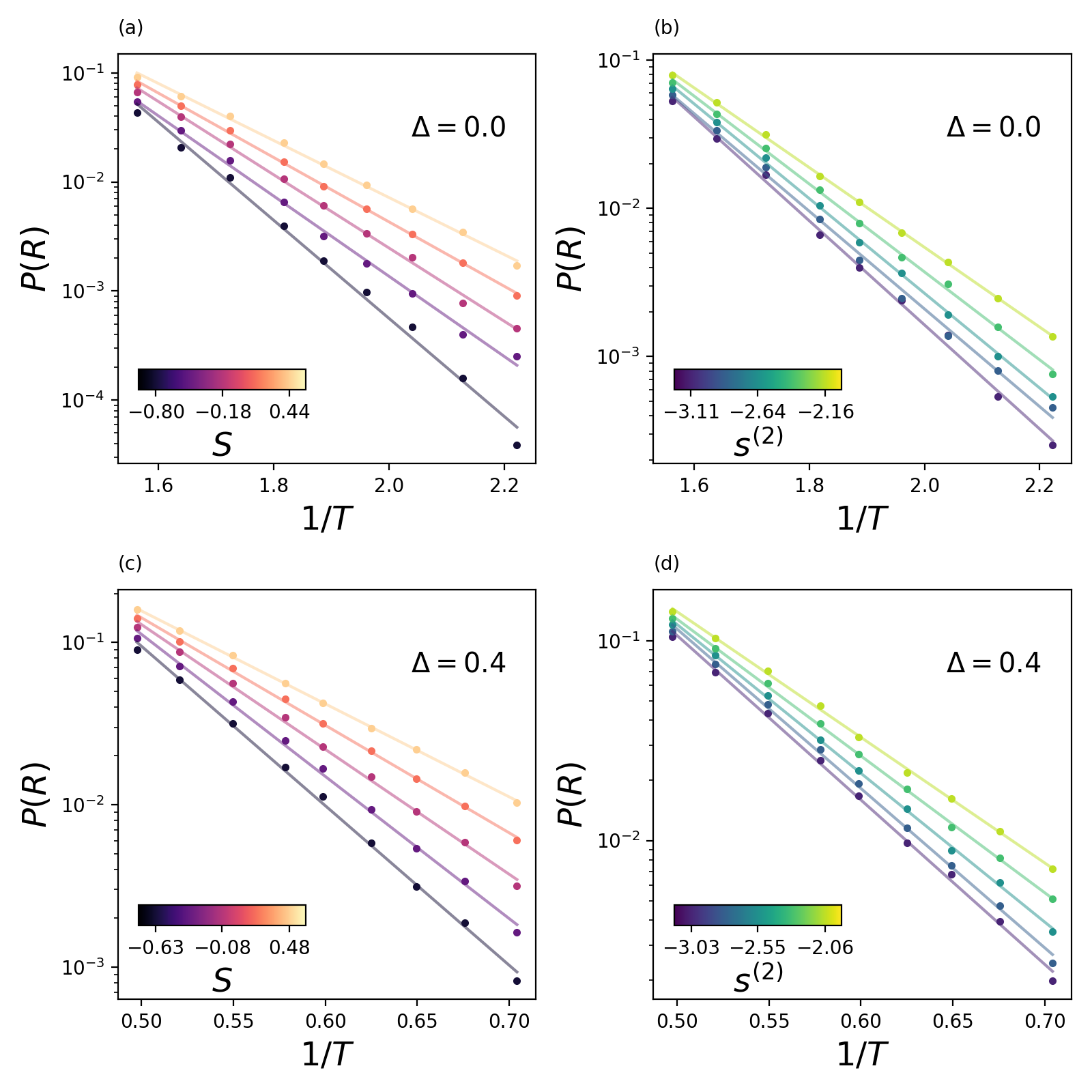}
    \caption{Probability of rearrangement against inverse temperature across the systems analyzed. Each system is separated into 5 equally spaced bins of softness, and at each temperature point we utilize a cutoff in $p_{hop}$ to measure the occurrences of local rearrangements. In subplots (a) and (c) softness is used to group particles with similar dynamics, leading to Arrhenius behaviour within each strata. In (b) and (d) we perform the same analysis, but instead using the local excess entropy as our structural proxy for dynamics. Both quantities are able to reproduce Arrhenius behaviour within the strata, though softness is more effective in capturing the range of dynamical behaviour within the system.}
    \label{Fig:ProbRearrang}
\end{figure}

Following previous analyses of softness, an Arrhenius-like relationship can be found as a function of temperature within the supercooled regime when tracking particles at constant $S$ across temperature. In Figure \ref{Fig:ProbRearrang}a,c we show this stratification for the probability of rearrangement in two of our systems binned by \textit{softness}. In each system, we take the 5th and 95th percentile bounds of \textit{softness}, and then use five equally spaced bins to aggregate our populations. Furthermore, these Arrhenius fits can be interpreted as relating to an Eyring-like equation $P_R(S)=\exp(\Sigma(S) - \Delta E(S) / T)$, where $\Sigma$ is the activation entropy, $\Delta E$ is the activation energy, and $T$ is the simulation temperature. When viewed this way, effective energetic and entropic barriers can be extracted from the fits, providing a connection between glassy local structures and the characteristic energy barriers governing the system's dynamics.

While these results regarding \textit{softness} are promising, one may wonder if the qualities that we find surprising here (the Arrhenius-like behavior and separation of relaxation rates) are hard to find using other measures of local packing. To explore this issue, we perform a similar analysis using our locally-defined excess entropy, as shown in Figure \ref{Fig:ProbRearrang}b,d. Although the quantity $s^{(2)}$ is not designed specifically to identify rearrangements, one still finds that the local excess entropy exhibits properties similar to those of $S$. For example, one still obtain fairly good Arrhenius fits to the strata with $s^{(2)}$, though the separation of these strata is not as large as with $S$. Depending upon the system, we observe at most an order of magnitude spread in the probability of rearrangement when looking at the coldest temperature sample when applying excess entropy. \textit{Softness}, on the other hand, is capable of extracting an additional order of magnitude or more in this spread.

\begin{figure}[H]
    \centering
    \includegraphics[width=1.0\textwidth]{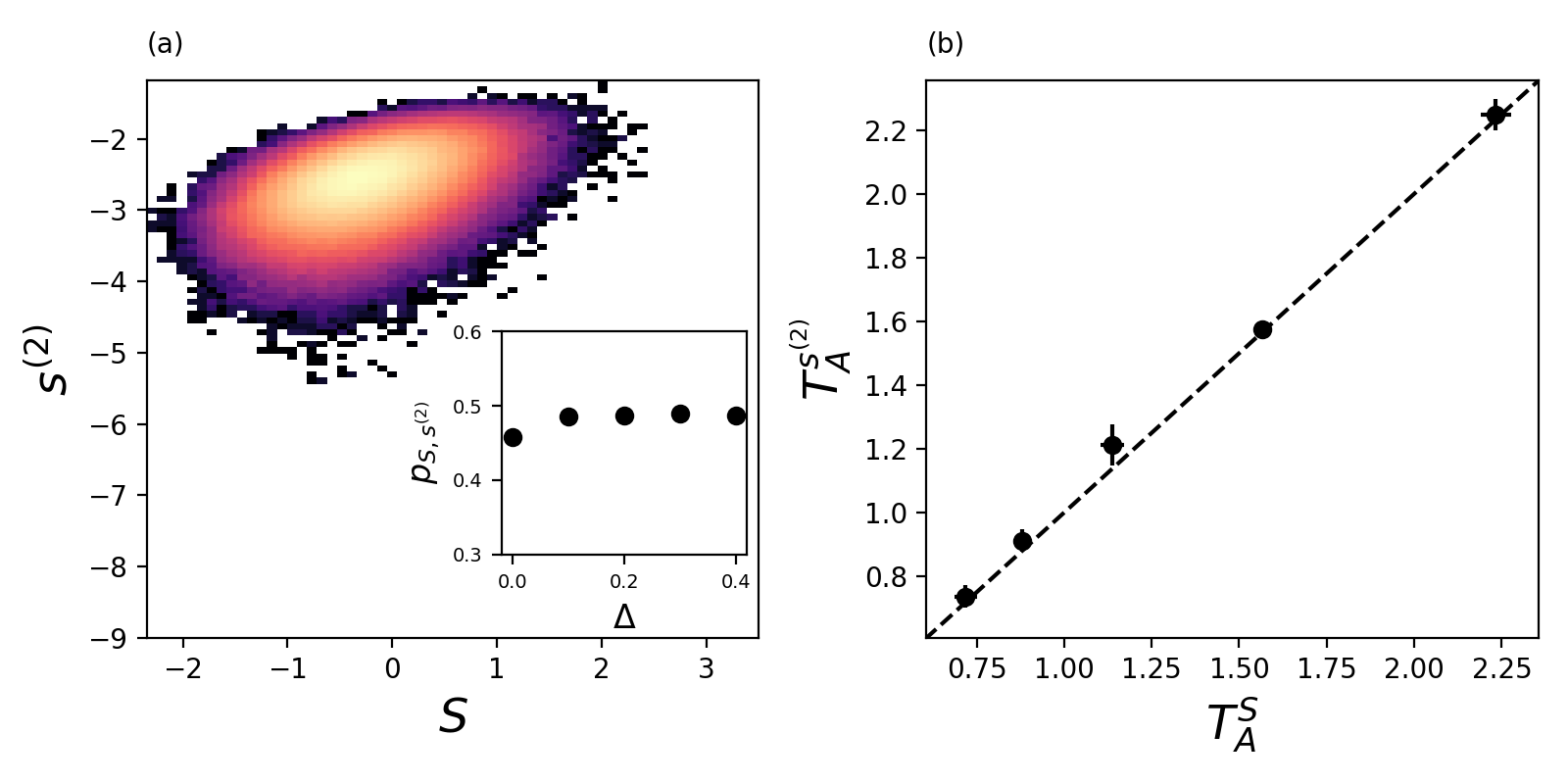}
    \caption{(a) 2D histogram of softness and local excess entropy at a particle level for the $\Delta=0.0$ system at $k_B T = 0.47$. (inset) Pearson correlation as a function of $\Delta$. We find a moderate correlation across the systems, which upon closer examination improves with increasing $\Delta$, likely due to the sharpening of the underlying RDF that enters the calculation. (b) Cross-over temperatures estimated from the Arrhenius fits in the supercooled regime using softness and excess entropy. A simple IQR outlier rejection method is applied to remove poor estimations that occur due to nearly co-linear Arrhenius fits. Good agreement is found between the estimates using the two quantities. }
    \label{Fig:QuantCorrelation}
\end{figure}

Next, we directly assess whether excess entropy and softness are correlated with each other. We do so by calculating the Pearson correlation coefficient $\rho_{S,s^{(2)}} = \tfrac{cov(S, s^{(2)})}{\sigma_S \sigma_{s^{(2)}}}$ between $S$ and $s^{(2)}$ as a function of the parameter $\Delta$. Across our systems we find a moderate correlation, as shown in the inset to Figure \ref{Fig:QuantCorrelation}a. We find an increase in the correlation between softness and excess entropy as $\Delta$ increases, which appears to plateau at $\approx 0.5$ (\ref{Fig:QuantCorrelation}a, inset). We believe that this is because the tighter well width is increasing the importance of structural entropy on dynamics, though this effect is limited. Our local excess entropy is successful in picking out some important features of softness, but ultimately falls short. We can understand this through the difference in the construction of these quantities. Excess entropy is directly penalizing fluctuations of the RDF relative to that of the ideal gas, i.e. $g(r)=1.0$. What seems to be crucial to softness' power is it's knowledge of the potential which it encodes transitively through the average $g(r)$ of a system and local fluctuations around $g(r)$ observed during the training process. 

Past studies were able to use dynamics (through dynamical heterogeneities) or structure (through \textit{softness}) to estimate the crossover temperature into the supercooled regime\citep{keys_excitations_2011, schoenholz_structural_2016} where the dynamics first become non-Arrhenius. In Figure \ref{Fig:QuantCorrelation}b we show the crossover temperatures estimated from the Arrhenius fits of softness and local excess entropy. Surprisingly, the crossover temperatures obtained through the two quantities agree relatively well. This is interesting to note, since traditional methods of extracting a crossover temperature require numerous measurements of long time dynamical quantities such as the diffusivity, viscosity, or alpha relaxation time. The distinct advantage that we hold by using excess entropy is that it requires no pre-training on data. Thus local excess entropy appears to be good and useful structural quantity to determine when the supercooled regime has been entered. We believe that estimating cross-over temperature is possible from $s^{(2)}$ because the structure captured by excess entropy is correlated with short-time dynamical heterogeneities in this regime.

One important observation here is that the Arrhenius trend found by softness may not be as unique as previous studies suggest. Instead, the observance of the Arrhenius trend in the supercooled regime may simply hinge on the quantity being at most weakly correlated with rearrangement probability. This also raises concerns about how to validate the performance of softness in general. Within thermal simulations, we have the freedom to use techniques such as the isoconfigurational ensemble and score our structural quantities by their correlation with propensity or other dynamical quantities averaged over the ensemble \citep{widmer-cooper_predicting_2006, bapst_unveiling_2020}. But in athermal sheared systems, where the rearrangement dynamics are binary in nature, it is difficult to frame softness' effectiveness as a correlation to average per-particle dynamics.

To summarize this section, we find that softness and local excess entropy, although moderately correlated, possess clear differences in their performance that affect their application as local indicators of rearrangement. Softness is clearly more suited to this task, but the moderate correlation suggests that there may be an underlying correspondence between these quantities. It remains that in cases where it is difficult or impossible to compute softness, the local excess entropy may make a fair proxy of dynamical behaviour and as a means to estimate the crossover temperature.

\subsection{Building ensembles with softness}

A broad distribution of relaxation rates is a hallmark of systems that exhibit glassy dynamics \citep{palmer_models_1984}. However, this feature greatly complicates the analysis of these systems with tools from statistical mechanics and thermodynamics. Thus, it is challenging to relate local structure to the relaxation rates of glassy systems, as is evident from the unsatisfying performance of purpose-built quantities discussed above. The concept of machine-learning \textit{softness} appears to be a fair solution to this issue in simple supercooled liquids, as recent works has demonstrated a clear connection between softness and relaxation rates in glassy systems\citep{schoenholz_structural_2016, schoenholz_relationship_2017}. This should be understood to be one of \textit{softness'} key contributions, that is, as a framework to infer dynamics from structure. And since these dynamics follow Arrhenius laws for a given softness, we can obtain information regarding local rearrangement barriers. But while softness possesses these valuable traits, it still cannot explain the dynamical behavior of systems in a way that connects to prior intuition.

\begin{figure}[H]
    \centering
    \includegraphics[width=1.0\textwidth]{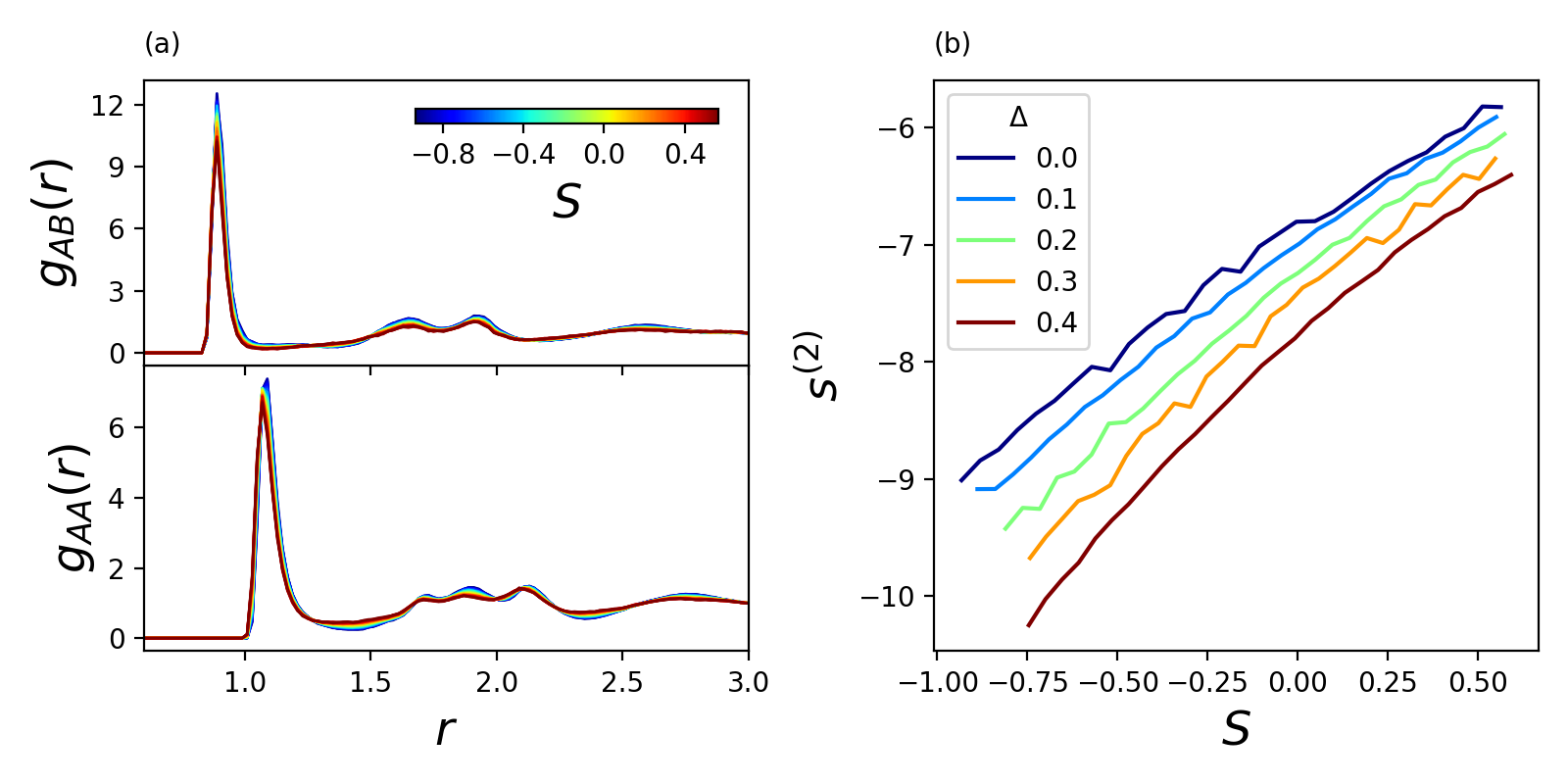}
    \caption{(a) Radial distribution functions of inherent structures of the $\Delta=0.0$, $k_B T = 0.49$ system when particles are grouped by softness. Lower-softness inherent structures tend to exhibit enhanced densities of particles at peaks in the RDF and suppression of those within the troughs. (b) Relationship between excess entropy and softness computed through RDFs of particle groupings with a given softness. The resulting relationship is not quite linear, but monotonically increasing.}
    \label{Fig:EntropyRDF}
\end{figure}

In order gain a deeper insight into the connection between thermodynamics and structure in glassy systems, we propose leveraging the predictive power of machine-learned indicators, like \emph{softness}, to partition our systems into sub-groupings that posses similar structure and dynamics. This can be done by simply grouping particles by their softness. Once these groupings are made, we posit that it is appropriate to compute thermodynamic quantities, like the excess entropy, in the traditional way by averaging over configurations, and that the results of these computations are qualitatively relatable to the actual dynamics. In this way, we propose thinking about the supercooled liquid not as possessing a homogeneous entropy, but as a number of distinct liquid states characterized by a spatially varying entropy, similar to the idea of ``entropic droplets'' from the random first-order transition theory \citep{kirkpatrick_stable_1987}.

\begin{figure}[H]
    \centering
    \includegraphics[width=1.0\textwidth]{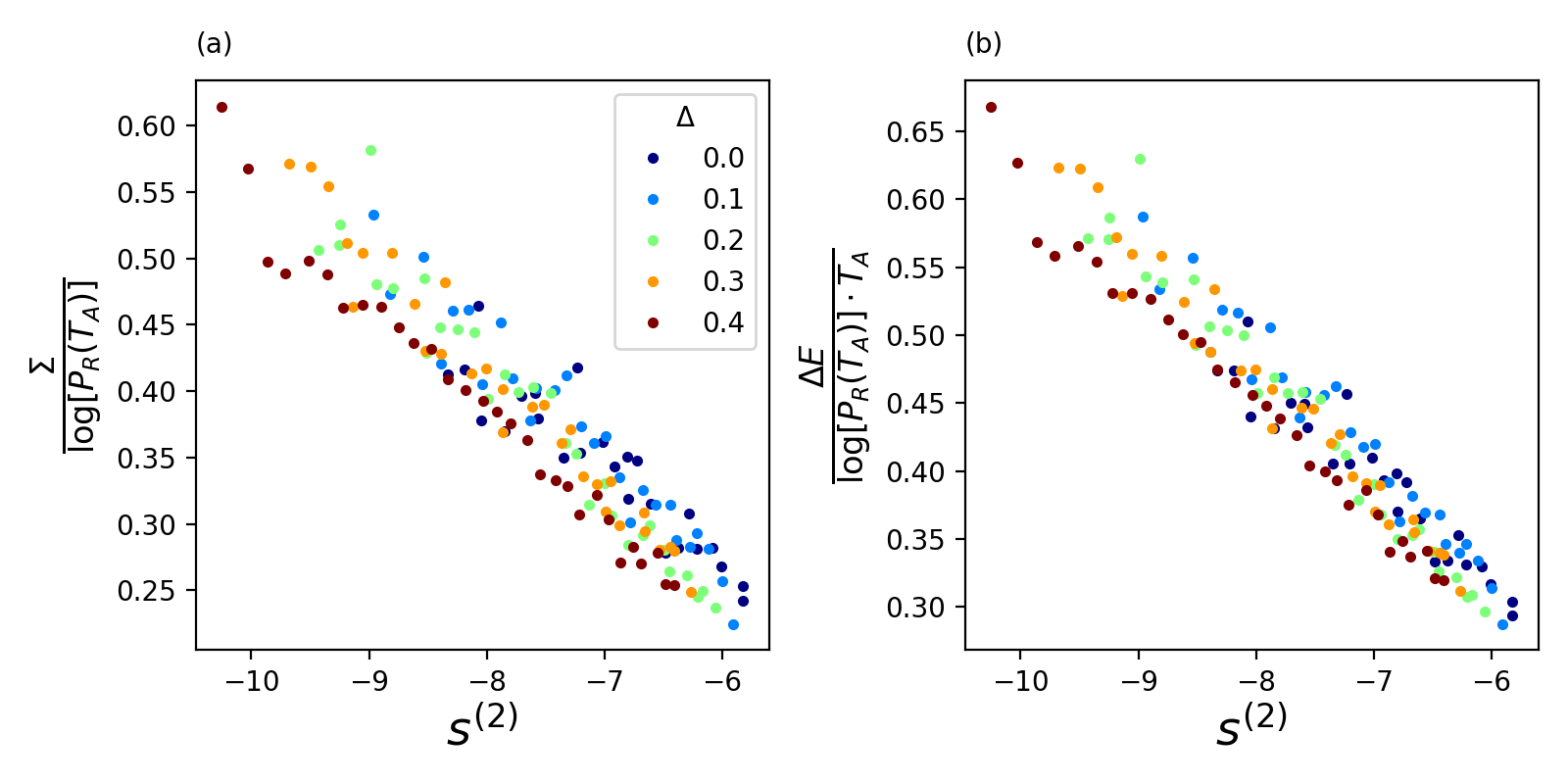}
    \caption{(a) Entropic and (b) energetic barriers as a function of excess entropy. We select similar temperatures relative to onset in each system, though the softness-grouped excess entropies of the inherent structure quenched states is insensitive to temperature. We find a simple rescaling by $\log[P_R(T_A)]$ and $T_A$ is sufficient to collapse the trends.}
    \label{Fig:Barriers}
\end{figure}

After grouping particles by softness, it is straightforward to construct the radial distribution functions from the inherent structure states. In Figure \ref{Fig:EntropyRDF}a we show examples of the A-A and A-B radial distribution functions (RDFs) calculated for the inherent structure configurations for particles of different softness. Particles of lower softness possess expected features, with neighbors more highly concentrated at peaks and deficient at troughs. We find that the inherent structure RDFs for the softness groupings are insensitive to the choice of temperature in the supercooled regime. These RDFs are used to compute $S$ and $s^{(2)}$ and extract a trend between softness and excess entropy of the particles binned by softness, as shown in Figure \ref{Fig:EntropyRDF}b. Using the excess entropy at a given softness from Figure \ref{Fig:EntropyRDF}b and the energy/entropy barriers extracted from Figures \ref{Fig:ProbRearrang}a and \ref{Fig:ProbRearrang}c, we can look for a relationship between the excess entropy and the barriers impeding rearrangements; we show this comparison in Figure \ref{Fig:Barriers}. Surprisingly, we find that the local entropic barriers of the systems, rescaled by the log-probability of rearrangement at $T_A$ ($\log[P_R(T_A)]$), appear to collapse the data as a function of the pair excess entropy. Interestingly, this relationship here appears linear, in contrast to the relationship between softness and rearrangement barriers, which often posses a slight quadratic character.

Figure \ref{Fig:Barriers}b shows a collapse in the energetic barriers by additionally rescaling by the onset temperature $T_A$ found in each system. Although this is somewhat unexpected, it follows from the fact that 1) we observe a collapse in the entropic barriers and 2) that the softness strata approximately intersect at the onset temperature $T_A$. Though a little algebra we find that
\begin{equation}
P_R(T_A) = \exp\left(\Sigma - \frac{\Delta E}{T_A}\right)
\end{equation}
\begin{equation}
\frac{\Delta E}{\log[P_R(T_A)] T_A} = \frac{\Sigma}{\log[P_R(T_A)]} - 1
\end{equation}
The required scaling with $\log[P_R(T_A)]$ appears to be an effect of our universal criterion for rearrangements in these systems, $p_{hop} > 0.2$. The results here are curious. The microscopic dynamics of the supercooled liquid, a system that has long been difficult to analyze using the tools of statistical mechanics and thermodynamics, appears to be well described by a thermodynamic structural measure, excess entropy, over these softness-grouped ensembles. Even if we take a leap and interpret $s^{(2)}$ here as a microscopic measure of configurational entropy in these systems, it is surprising that this also well predicts the barrier heights as the potential is modified. The only other critical information to forming the observed collapse is the onset temperature, which is straightforward to measure.

\subsection{Discussion}

Our first set of results suggest that quantities such as the local excess entropy are able to recover many of the notable features attributed to softness, e.g., Arrhenius trends in the supercooled regime and an ability to infer the onset temperature for supercooled dynamics. While the overall performance (in terms of predicting particle rearrangements) of excess entropy is not as robust as softness, it does surprisingly well for a quantity that has not been aggressively optimized. Though the simple application of excess entropy as a local, non-averaged estimate is indeed somewhat naive, and information of the interactions must be somehow incorporated (even indirectly) to elevate the local estimate of excess entropy to at least the level of softness. A more intricate exploration of the transition states may help to reveal these connections between the local inherent structures and the barriers to rearrangement. Particularly, if one considers ensembles of transition states with the same energy barrier height, what average behavior emerges for the per-particle energies with distance? Is it possible that we could intuit this from a combination of $U(r)$ and its derivatives?

Furthermore, our analysis using softness-grouped ensembles exposes a remarkably simple connection between excess entropy and the rearrangement barriers within supercooled liquids. Even though the supercooled state is fraught with many complications due to its rich behavior and metastable nature, there appears to be an optimal partitioning of particles that allows us to make connections to equilibrium thermodynamics. Unfortunately, however, effectively obtaining these groups at the moment requires either dynamical sampling techniques like isoconfigurational ensembles or machine-learned quantities that pose superior correlations with rearrangement dynamics. Quantities derived from low-frequency modes may also be good candidates for this type of aggregation \citep{tong_order_2014}. The framework presented here is a possible path for framing thermodynamics in supercooled liquids, and it may help us to better understand the physical principles responsible for the emergence of glassy behavior at the onset temperature.

\subsection{Acknowledgements}
This work is funded by University of Pennsylvania’s MRSEC NSF-DMR-1720530.

\bibliographystyle{plainnat}
\bibliography{main}

\end{document}